\documentclass[twocolumn]{aastex62}
\usepackage{graphicx}	
\usepackage[fleqn]{amsmath}	
\usepackage{eqnarray}	
\usepackage{amssymb}	
\usepackage{bm}		    
\usepackage[T1]{fontenc}
\usepackage{natbib}
\usepackage{listings}
\usepackage{multirow}
\submitjournal{ApJ}

\shorttitle{Lost in space}
\shortauthors{Regály, Fröhlich \& Vinkó}
\graphicspath{{./}{figures/}}

\begin{document}

\title{\large{\sc{Lost in space: companions' fatal dance around massive dying stars}}}

\correspondingauthor{Zsolt Regály}
\email{regaly@konkoly.hu}

\author[0000-0001-5573-8190]{Zsolt Reg\'aly}
\affiliation{Konkoly Observatory, Research Centre for Astronomy and Earth Science, Konkoly-Thege Mikl\'os 15-17, H-1121, Budapest, Hungary}
\affiliation{CSFK, MTA Centre of Excellence, Konkoly Thege Miklós út 15-17., H-1121, Budapest, Hungary}

\author[0000-0003-3780-7185]{Viktória Fröhlich}
\affiliation{Konkoly Observatory, Research Centre for Astronomy and Earth Science, Konkoly-Thege Mikl\'os 15-17, H-1121, Budapest, Hungary}
\affiliation{CSFK, MTA Centre of Excellence, Konkoly Thege Miklós út 15-17., H-1121, Budapest, Hungary}
\affiliation{E\"otv\"os Lor\'and University, P\'azm\'any P\'eter s\'et\'any 1/A, H-1117 Budapest, Hungary}

\author[0000-0001-8764-7832]{József Vinkó}
\affiliation{Konkoly Observatory, Research Centre for Astronomy and Earth Science, Konkoly-Thege Mikl\'os 15-17, H-1121, Budapest, Hungary}
\affiliation{CSFK, MTA Centre of Excellence, Konkoly Thege Miklós út 15-17., H-1121,  Budapest, Hungary}
\affiliation{ELTE E\"otv\"os Lor\'and University, Institute of Physics, P\'azm\'any P\'eter s\'et\'any 1/A, Budapest, 1117 Hungary}
\affiliation{Institute of Physics, University of Szeged, D\'om t\'er 9, Szeged, H-6720, Hungary}




\begin{abstract}

Discoveries of planet and stellar remnant hosting pulsars challenge our understanding, as the violent supernova explosion that forms the pulsar presumably destabilizes the system.
Type II supernova explosions lead to the formation of eccentric bound systems, free-floating planets, neutron stars, pulsars, and white dwarfs.
Analytical and numerical studies of high mass-loss rate systems based on perturbation theory so far have focused mainly on planet-star systems.
In this paper, we extend our understanding of the fate of planet-star and binary systems by assuming a homologous envelope expansion model using a plausible ejection velocity ($1000-10,000\,\mathrm{km\,s^{-1}}$), envelope and neutron star masses.
The investigation covers secondary masses of $1-10\,M_\mathrm{J}$ for planetary companions, and $1-20\,M_\odot$ for stellar companions.
We conduct and analyze over 2.5 million simulations assuming different semi-major axes ($2.23 - 100\,\mathrm{au}$), eccentricities ($0-0.8$), and true anomalies ($0-2\pi$) for the companion.
In a homologous expansion scenario, we confirm that the most probable outcome of the explosion is the destabilization of the system, while the retention of a bound system requires a highly eccentric primordial orbit.
In general, a higher ejecta velocity results in a lower eccentricity orbit independent of secondary mass.
The explanation of close-in pulsar planets requires exotic formation scenarios, rather than survival through the type II supernova explosion model.
Post-explosion bound star systems gain a peculiar velocity (<100\,km/s), even though the explosion model is symmetric.
The applied numerical model allows us to derive velocity components for dissociating systems.
The peculiar velocities of free-floating planets and stellar corpses are in the range of $10^{-6}-275\,\mathrm{\,km\,s^{-1}}$.

\end{abstract}



\section{Introduction}
\label{sec:intro}

As of today, the majority of confirmed exoplanets orbit around main-sequence stars.
However, with the advancement of detection methods, the number of confirmed planets orbiting around pulsars now totals up to eight, including
the first confirmed exoplanetary system, PSR B1257+12 (\citealp{wolszczan-frail, wolszczan-1994, wolszczan-2012}), and recently discovered PSR B1620-26 \citep{backer-etal, thorsett-etal, sigurdsson-etal}, PSR J1719-1438 \citep{bailes-etal-2011}, PSR B0943+10 \citep{suleymanova-rodin}, PSR B0329+54 \citep{starovoit-rodin}, and PSR J2322-2650 \citep{spiewak-etal}.

For pulsar planets, there are three possible formation mechanisms: surviving the supernova in a stable orbit, planet capture, and hypothetical second-generation formation.
The existence of planets and especially planetary systems around stellar remnants challenges our understanding. 
In the first scenario, the survival of a planet in a stable orbit is questionable, since neutron stars (NSs) form in violent type II supernova explosions (SN~II; see, e.g., \citealp{podsiadlowski, veras-etal, fagginger-portegies}).
Moreover, in this case, a massive planet-hosting progenitor is required, at least $8M_\odot$, which is much higher than the largest planet-hosting star observed today. 
Note, however, that planets orbiting massive stars are also hard to detect \citep{kennedy-kenyon, han-etal}.

The second scenario implicitly assumes that a pulsar can capture a free-floating planet (or even multiple planets subsequently, in the case of PSR B1257+12).
This is presumably a highly improbable event \citep{podsiadlowski}, even though the number of observed free floaters in many different environments constantly increases (see the surveys of \citealp{lucas-roche, zapatero-osorio-etal, bihain-etal, sumi-etal, miret-roig-etal}).

In the third scenario, planets can form in the circumstellar fallback material around the SN\,II (\citealp{colgate, chevalier, bailes-etal-1991, phinney-hansen, podsiadlowski, lohmer-etal}, numerical study in \citealp{perna-etal}).
Planetary debris disks around pulsars might form after the destruction of the planet after the SN~II; see \citet{Veras2016} for a summary.
These processes have already been investigated by \citet{perets} in general, around post-AGB binaries by \citet{kluska-etal} and in post common envelope evolution by \citet{bear-soker}, \citet{schleicher-dreizler}, \citet{schleicher-etal} and \citet{kostov-etal}.
Other, more exotic theories, e.g., the tidal disruption or evaporation of a companion star, have been discussed in \citet{martin-etal}.

Due to the large amount of mass loss suffered by the planet-hosting star during an SN\,II event, orbital stability breaks down.
Mass loss in a two-body problem (with both bodies considered to be point masses) has been studied for well over a century, first by \citet{gylden} and \citet{mestschersky}.
More general solutions were then developed (see \citealp{razbitnaya} for a summary), and the problem was also solved for binary star systems by \citet{jeans}. 
The perturbed two-body problem is thoroughly investigated in \cite{burns}.
The perturbation equations of the two-body system, describing the secular variations in orbital elements are given in \citet{hadjidemetriou-1963} and \citet{verhulst}.

Numerical solutions were pioneered by \citet{hadjidemetriou-1963} and \citet{hadjidemetriou-1966-a}. 
In these studies, the mass-loss rate was assumed to be low; thus, secular variations were only observed in the semi-major axis, while the eccentricity remained secularly constant.
\citet{hadjidemetriou-1966-b} showed that the orbital eccentricity of the secondary may vary due to extreme mass loss in a supernova explosion.
This may cause the eccentricity to grow above unity; thus, the secondary breaks away from its host star and becomes a free-floating object.

Numerical studies of the variable-mass two-body problem have so far been adiabatic, where the secondary's eccentricity is thought to remain constant throughout the simulations \citep{debes-sigurdsson}.
The pioneering work of \citet{veras-etal} provided both analytical and numerical solutions to the perturbation equations for mass-loss regimes resembling an SN II phenomenon. 
That investigation adopted a simplified, constant mass-loss rate prescription of \citet{hurley-etal}.
The aim was to compare analytical and numerical calculations, assuming a planetary-mass secondary and a wide range of primary masses.
The major conclusion was that planets orbiting neutron stars born in an SN~II explosion remain bound only in limited cases, and for a wide range of initial conditions the planet gains a large eccentricity.

In this study, we extend upon the works of \citet{veras-etal} by the means of numerical simulations of mass loss in planet-star systems, as well as in binary star systems.
For mass loss, we employ a homologous expansion model, which presumably better describes the mass loss occurring during envelope expansion.
We give a detailed analysis of orbital elements and component velocities of over two and a half million numerical simulations.
We investigate the possible outcomes of SN~II explosions, in which bound planetary and binary star systems, as well as free-floating planets, stars, and pulsars, form.

This paper is structured as follows:
the numerical model applied to this study and the investigated parameter space are described in Section \ref{sec:model}.
Our results and numerical analysis are presented in Section \ref{sec:results}.
Section \ref{sec:discussion} deals with the discussion of the fate of SN~II systems.
The paper closes with our conclusions and remarks on future investigations in Section \ref{sec:conclusions}.

\section{Numerical model}
\label{sec:model}

\subsection{Numerical integrator}
\label{sub:int}

To model the change of the orbital parameters of the two-body system (central exploding star and a companion, which is assumed to be either a stellar companion or a planet), we solve the equations of motion of the system numerically in two dimensions.
This way, unlike the analytic methods presented in \citet{veras-etal}, we can examine a secondary whose mass is nonnegligible compared to that of the primary. 
The barycenter is not fixed on the primary, and thus we can analyze the SN\,II events in a binary star system.
The equations of motion for the primary and secondary components are
\begin{equation}
    \ddot{\textbf{r}}'_\mathrm{pri}=-M_\mathrm{sec} \ G \ \frac{\textbf{r}'_\mathrm{pri}}{r_\mathrm{pri}^{'3}},
    \, \,  \,  \, \, \, \, \, \, \, \, 
    \ddot{\textbf{r}}'_\mathrm{sec}=-M_\mathrm{in} \ G \ \frac{\textbf{r}'_\mathrm{sec}}{r_\mathrm{sec}^{'3}},
\end{equation}
where $G$ is the gravitational constant, and $\textbf{r}'_\mathrm{pri}$ and $\textbf{r}'_\mathrm{sec}$ are the position vector of the primary and secondary components in the inertial frame, respectively. 
$M_\mathrm{sec}$ is the secondary's mass, which is a constant value, while the mass inside the secondary's orbit, $M_\mathrm{in}$, changes due to the envelope loss of the primary.
To calculate $M_\mathrm{in}$, we develop a homologous envelope loss model; see details in the next section.

Working in Python 3.8.8 and opting for SciPy's \verb|integrate.solve_ivp()| function with an explicit Runge--Kutta method of order 8 (DOP853), we integrate the primary and the secondary's barycentric position and velocity vectors through the explosion and beyond.
The method works with an adaptive step size to reach the desired precision.
$M_\mathrm{in}$ is updated at every time step during numerical integration.

\subsection{Homologous expansion model}
\label{sub:homologous}

When a massive star's life ends in a type II supernova explosion, the stellar core collapses and forms a neutron star.
Meanwhile, the envelope falls onto this dense core and bounces back.
Through this process, a shock wave is initiated, which is then responsible for the ejection of the stellar outer shell.
The shock wave propagates through the envelope and reaches the stellar surface. 
The mass of the ejecta could be as high as 95\% of the initial stellar mass \citep{smartt-etal}.
Here, we assume that the envelope expands in a spherically symmetric, homologous fashion 
\citep[e.g.][]{arnett80, bw17}, 
which results in the following: (1) the velocity of each ejected layer is a linear function of distance; (2) the velocity of the outermost layer is independent of time; and (3) the density profile of the ejected material is also time-invariant.

The assumed spherical geometry of the expanding ejecta, at least during the first $\sim 100$ days after explosion, is consistent with spectropolarimetric observations of hydrogen-rich (type II) core-collapse supernovae \citep[e.g.][]{bw17, nagao21}. The observed continuum polarization of the majority of such supernovae is found to be close to zero shortly after explosion, but tends to increase as the ejecta expands. This suggests that the inner parts of the ejecta may deviate from spherical symmetry, even though interaction with the surrounding circumstellar matter (CSM) may also contribute to the observed polarization. Motivated by the initial spherical symmetry of the ejecta as observed (and admitting that an asymmetric geometric configuration would introduce additional complexity in our calculations), we assume spherical symmetry of the expanding supernova ejecta in our calculations.

In the following, we summarize the homologous envelope expansion model applied, e.g., in \citet{vinko04}. 
Due to the homologous expansion, the velocity of the envelope at a given distance $r(t)$ from the center is $v(r,t) = \left(r(t)/R(t) \right) v_\mathrm{max}$, where $R(t)$ is the time-dependent radius of the outermost layer of the expanding ejecta and $v_\mathrm{max}$ is the maximum expansion velocity at $R$.
Throughout the expansion, $v_\mathrm{max}$ is assumed to be constant.
Initially the envelope radius $R = R_0$ is equal to the stellar radius, and is assumed to be $R_0 = 500\,R_\odot$ ($\sim 2.23$\,au) throughout this study.
We assume that the inner 10\% of the star contains a constant density core, which has a fractional radius $r_\mathrm{c}=0.1R$.

Introducing the relative (comoving) distance coordinate as $x=r(t)/R(t)$, the envelope velocity at $x$ is $v(x,t)=x \cdot v_\mathrm{max}$.
Using the above equation and defining $\Delta t$ as the time elapsed since shock breakout, the distance of a given mass layer from the SN center can be expressed as 
\begin{equation}
    r(t) = r(0) + v(r,t) \Delta t = x(R_\mathrm{0}+v_\mathrm{max} \Delta t).
    \label{eq:envelope_r}
\end{equation}
When this layer reaches the orbit of the secondary body, $r = r_\mathrm{sec}$, thus 
\begin{equation}
    x(r_\mathrm{sec}) \equiv x_\mathrm{sec}=\frac{r_\mathrm{sec}}{R_\mathrm{0}+v_\mathrm{max} \Delta t}.
    \label{eq:x_p}
\end{equation}

\begin{figure}
    \centering
    \includegraphics[width=\columnwidth]{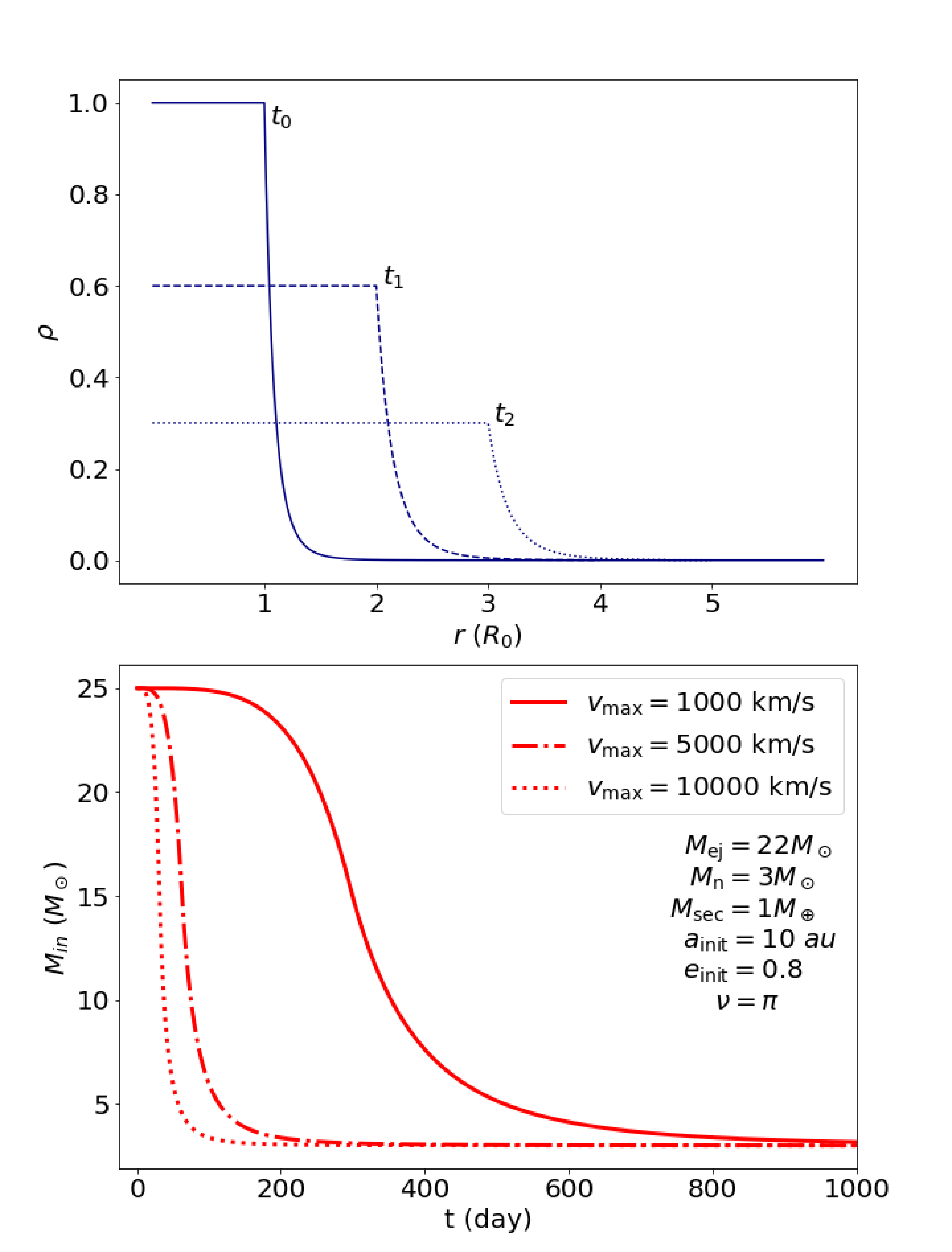}
    \caption{Upper panel: radial density profile of the core and envelope at three phases $t_0<t_1<t_2$ of SN\,II explosion. Density is normalized with the initial core density. Lower panel: time evolution of the total mass residing inside the secondary's orbit for expansion velocities 1000, 5000, and $10,000\,\mathrm{km\,s^{-1}}$.}
    \label{fig:density}
\end{figure}

To determine the mass inside $r_\mathrm{sec}$, $M_\mathrm{in}$, three cases should be distinguished: 
$x_\mathrm{sec}>1$, or either $x_\mathrm{sec} > x_\mathrm{c}$ or $x_\mathrm{sec}<x_\mathrm{c}$.
In the first case, all of the stellar mass resides inside the secondary's orbit, and therefore $M_\mathrm{in}=M_\mathrm{ej}+M_\mathrm{n}$, where $M_\mathrm{ej}$ is the mass of the stellar ejecta, and $M_\mathrm{n}$ the mass of the remnant neutron star.
In the latter cases, we take into account the change of envelope density due to homologous expansion. 
The density profile of the core is considered to be spatially constant ($\rho_0$), while the density of the envelope follows a power law, 
\begin{equation}
    \rho=\rho_0 \left(\frac{x}{x_\mathrm{c}}\right)^{-n},
    \label{eq:density}
\end{equation}
where $n=7$ is assumed if $x > x_\mathrm{c}$, and $n=0$ otherwise.
The core density function changes in time according to
\begin{equation}
    \rho_0(t) = \frac{3 M_\mathrm{c}}{4 \pi} \left ( x_\mathrm{c} (R_\mathrm{0}+v_\mathrm{max} \Delta t) \right )^{-3}.
    \label{eq:coredens}
\end{equation}
If $r_\mathrm{sec} > r_\mathrm{c}$ then the mass residing within the secondary's orbit is
\begin{equation}
    M_\mathrm{in}= M_\mathrm{n} + M_\mathrm{c} + \int_{r_\mathrm{c}}^{r_\mathrm{sec}} 4 \pi r^2 \rho(r) dr,
    \label{eq:min_def}
\end{equation}
where $r_c$ is the radius of the core.
If $r_\mathrm{sec}>r_\mathrm{c}$, the above equation, after some trivial algebra, gives
\begin{align}
    \begin{split}
        M_\mathrm{in}&=M_\mathrm{n} + M_\mathrm{c}+ 4 \pi \rho_\mathrm{0} R_\mathrm{0}^3 x_\mathrm{c}^n \int_{x_\mathrm{c}}^{x_\mathrm{sec}} x^{2-n} dx =\\
        &= M_\mathrm{n} + M_\mathrm{c} \left [ 1+ \frac{3}{n-3} \left ( 1- \left ( \frac{x_\mathrm{sec}}{x_\mathrm{c}} \right )^{3-n} \right ) \right ].    
    \end{split}
    \label{eq:min_xp>xc}
\end{align}
That said, if $x_\mathrm{sec}<x_\mathrm{c}$,
\begin{align}
    \begin{split}
    M_\mathrm{in}&=M_\mathrm{n}+ 4 \pi \rho_\mathrm{0} R_\mathrm{0}^3  \int_{0}^{x_\mathrm{sec}} x^2 dx =\\
    &= M_\mathrm{n} + M_\mathrm{c} \left ( \frac{x_\mathrm{sec}}{x_\mathrm{c}} \right )^3.
    \end{split}
    \label{eq:min_xp<xc}
\end{align}
With the assumed density profile, the mass of the ejecta, $M_\mathrm{ej}$ can be expressed as
\begin{align}
    \begin{split}
        M_\mathrm{ej} &= 4 \pi R_0^3 \rho_\mathrm{0} \left ( \int_{0}^{x_\mathrm{c}} x^2 dx + x_\mathrm{c}^n \int_{x_\mathrm{c}}^{1} x^{2-n} dx \right ) = \\
        &= 4 \pi R_0^3 \rho_\mathrm{0} \left ( \frac{x_\mathrm{c}^3}{3} + \frac{x_\mathrm{c}^n-x_\mathrm{c}^3}{3-n} \right ).
    \end{split}
    \label{eq:mej}
\end{align}
From this expression, one can derive the core density as
\begin{equation}
    \rho_\mathrm{0} = \frac{M_\mathrm{ej}}{4 \pi R_0^3} \left ( \frac{x_\mathrm{c}^3}{3} + \frac{x_\mathrm{c}^n-x_\mathrm{c}^3}{3-n} \right )^{-1}.
    \label{eq:rho0}
\end{equation}
Now the core mass can be calculated as
\begin{equation}
    M_\mathrm{c}=\frac{4 \pi R_0^3 x_\mathrm{c}^3}{3} \rho_\mathrm{0}= \frac{M_\mathrm{ej}} {1+ \frac{3}{n-3} \left( 1- x_\mathrm{c}^{n-3} \right ) }.
    \label{eq:mcore}
\end{equation}
The upper panel of Figure \ref{fig:density} shows an example of the evolution of the radial density profile of the stellar core and envelope at three different phases.
The lower panel of Figure \ref{fig:density} shows some examples of the time evolution of the total mass residing inside the secondary for three different expansion velocities.

\subsection{Models investigated}
\label{sub:param_space}

The total integration time for each specific model was 5 million days (about 13,700 yt).
On average, by the end of the simulations, the measured mass-loss rate is $~10^{-11} - 10^{-18} ~M_\odot\,\mathrm{day^{-1}}$, meaning that the mass loss saturates; see the lower panel of Figure~\ref{fig:density}.
This allows the monitored values (semi-major axis, $a$, eccentricity, $e$, the velocity of the primary, $v_\mathrm{pri}$, and the secondary, $v_\mathrm{sec}$) to relax by the end of the simulation.
Using relative and absolute tolerance values of $10^{-12}$ for a fixed-mass two-body problem with an assumption of a high-mass secondary ($10\,M_\odot$), the relative error in total energy is on the order of $10^{-9}$ by the end of integration time.

We monitor the perturbation and evolution of the secondary's eccentricity and semi-major axis throughout simulations using the following equations:
\begin{equation}
    e = \sqrt{1 + 2 h \left ( \frac{ \textbf{r}_\mathrm{sec} \times \textbf{v}_\mathrm{sec} } {\mu} \right )^2},
    \\\\
    a = - \frac{\mu}{2h},
    \label{eq:e,a}
\end{equation}
where $\textbf{r}_\mathrm{sec}$ and $\textbf{v}_\mathrm{sec}$ are the secondary's heliocentric position and velocity vectors, respectively.
The heliocentric position and velocity vectors of the primary are derived from the barycentric components.
In Equations \ref{eq:e,a}, the energy term reads
\begin{equation}
    h= \frac{1}{2} v_\mathrm{sec}^2 -\frac{\mu}{r_\mathrm{sec}},
\end{equation}
and $\mu=G(M_\mathrm{sec}+M_\mathrm{in})$.
Here, $r_\mathrm{sec}$ and $v_\mathrm{sec}$ are the secondary's heliocentric distance and speed, respectively.

To map a wide range of scenarios, we investigate different initial conditions.
Model set A maps various orbital elements (initial semi major axis $a_\mathrm{init}$, initial eccentricity $e_\mathrm{init}$ and initial true anomaly $\nu$) and secondary masses ($M_\mathrm{sec}$).
In this set, we also test different envelope and remnant masses.
Neutron star mass is constrained by the Chandrasekhar limit, and an upper limit of $3\,M_\odot$ is adopted based on theory and observational evidence (\citealp{kalogera-baym, clark-etal}). 
As for the velocity of the outermost envelope layer, i.e. the ejection velocity, we adopt the $1000-10,000\,\mathrm{km\,s^{-1}}$ range based on \citet{hamuy-pinto}.
In the investigated semi-major axis range this results in the ejecta reaching the secondary well before it completes its first orbit after the explosion.
The parameters used in Model set A can be found in Table \ref{tab:params}.
This gives us a total of 1,875,000 models to investigate.
Only those models are analyzed where the secondary mass is below that of the primary. 

To map the effect of true anomaly on the evolution of the system, we run Model set B, an additional 720,500 models.
The settings of these models are $M_\mathrm{ej}=6.5\,M_\odot$ and $M_\mathrm{n}=3\,M_\odot$.
A total of 1,441 different true anomaly values are sampled linearly from the $[0-2\pi]$ range, and 25 $a_\mathrm{init}$ values from the range $[R_0-100\,\mathrm{au}]$.
We assumed that the secondary orbits outside the envelope in each model, as whether planets can survive in stellar envelopes still remains a question (see, e.g., \citealp{setiawan-etal}, \citealp{bear-etal} and more recently \citealp{lagosetal}, \citealp{chamandyetal}, \citealp{szolgyenetal}).
We investigate both a planetary ($M_\mathrm{sec}=1\,M_\oplus$) and stellar ($M_\mathrm{sec}=1,\,3,\,5\, \text{\ and \ } 7\,M_\odot$) companions, whose eccentricity is $0.4$ or $0.8$.
The velocity of the envelope is set to be $1000 \ \mathrm{km/s}$ or $10,000 \ \mathrm{km\,s^{-1}}$.

A given model is considered to be bound if the secondary's eccentricity value remains under unity, and unbound if it grows beyond unity by the end of the simulation.
Models with a planetary and a stellar-mass companion show qualitatively different outcomes: bound planets, (unbound) free-floating planets, bound stars, and free-floating stars. 

\begin{table}[h!]
\begin{center}
\caption{Parameters Used in Model Set A}
\begin{tabular}{ll}
\hline
\hline
$a_\mathrm{init}$  & $[R_0 - 100 \ \mathrm{au}]^\dag$ \\ 
$e_\mathrm{init}$  & $0.0,0.1,0.2,0.4,0.8$  \\ 
$\nu$  & $0.0,\pi/4,\pi/2,3\pi/4,\pi$  \\ 
$v_\mathrm{max}$  & $[1000 - 10,000] \ \mathrm{kms^{-1\dag}}$  \\ 
$M_\mathrm{ej}$  & $6.5\,M_\odot, 10\,M_\odot, 15\,M_\odot, 22\,M_\odot$  \\ 
$M_\mathrm{n}$  & $1.5\,M_\odot, 2\,M_\odot, 3\,M_\odot$  \\ 
\multirow{2}{*}{$M_\mathrm{sec}$} & $1\,M_\oplus, 10\,M_\oplus, 1M_\mathrm{J}, 10M_\mathrm{J}, $   \\
                                                    & $1\,M_\odot, 2\,M_\odot, 3\,M_\odot, 5\,M_\odot, 10\,M_\odot, 20\,M_\odot$ \\ \hline
\end{tabular}
\end{center}
\label{tab:params}
\tablecomments{$M_\oplus$ denotes Earth, $M_\mathrm{J}$ Jupiter, and $M_\odot$ solar mass. $^\dag$Twenty-five values are sampled linearly from the given ranges.}
\end{table}

\section{Results}
\label{sec:results}

\subsection{Bound cases}
\label{sub:bound}

\begin{figure*}
    \centering
    \includegraphics[width=0.75\textwidth]{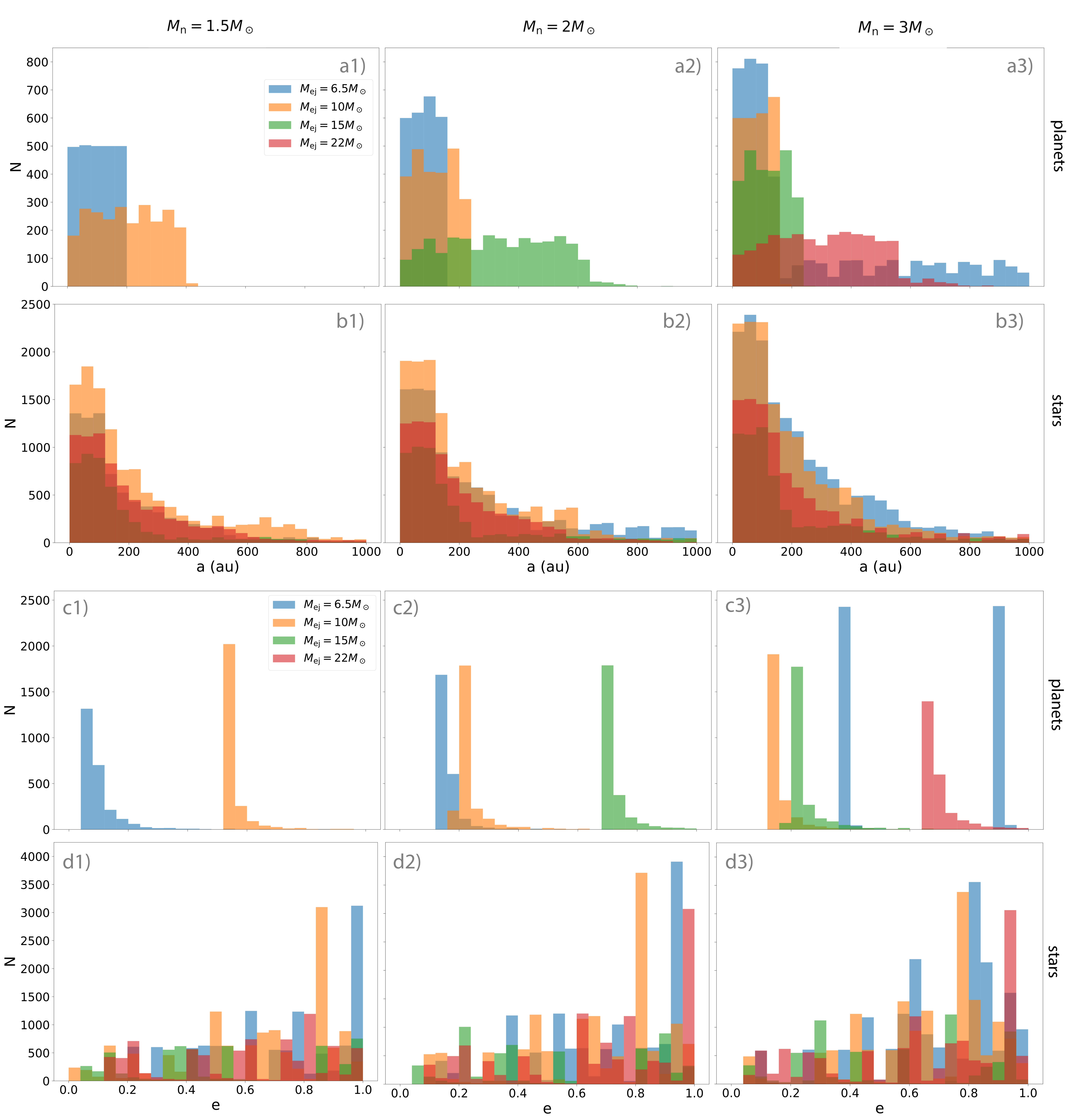}
    \caption{Histograms of the orbital parameters (top two rows: $a$, bottom two rows: $e$ for planets and stars, respectively) of bound systems at the end of integration. Panels show cases with different $M_\mathrm{n}$ values. Colors represent different $M_\mathrm{ej}$ values, also shown in the key.}
    \label{fig:hist_bound}
\end{figure*}

In this section, we investigate cases in which a bound system is formed.
In the center, there is always a neutron star, while the companion can be either a planet or a star, depending on the initial condition.
Figure \ref{fig:hist_bound} shows the orbital parameters ($a$ and $e$) of the companion planet or star for different initial conditions, calculated according to Equations\,(\ref{eq:e,a}) at the end of the integration when these orbital parameters are relaxed.

With regards to the incidence of a bound state, we find that by increasing the mass of the neutron star, the chance of forming a bound orbit also increases, as shown in panels a1--b3 of Figure \ref{fig:hist_bound}.
This finding is in agreement with \citet{veras-etal} for planetary companions and is also found to be valid for stellar companions according to our simulations.
The formation of a bound state requires a planet initially close to the apocenter (see details in Section \ref{sub:true_anomaly}.). 
We also find that the semi-major axis of bound planets increases with increasing the mass of the ejecta, which can be seen in panels a1--a3 of Figure\,\ref{fig:hist_bound}. 

By analyzing semi-major axis histograms normalized by $a_\mathrm{init}$, a clear trend can be identified in the planetary case.
The ratio of the final-to-initial semi-major axis values shows discrete peaks (due to the assumption of discrete ejecta mass values), and the value of the peak centers increases with ejecta mass.
Interestingly, there is also a minimum value of about 2 for this ratio.
By contrast, for stellar-mass companions, such a trend cannot be identified. 

The final eccentricities of bound planets, shown in panels c1--c3, are found to be clumpy and the number of clumps increases with increasing $M_\mathrm{n}$.
This is due to the fact that we use discrete $M_\mathrm{ej}$ and $M_\mathrm{n}$ values, and the number of discrete $M_\mathrm{ej}/M_\mathrm{n}$ bound models increases with $M_\mathrm{n}$.
There is an eccentricity minimum (being the most probable value), for all pairings, that also decreases with $M_\mathrm{n}$ and increases with $M_\mathrm{ej}$.
Final $a$ and $e$ values are independent of planetary mass.

We emphasize that the initial eccentricity of a planetary companion should be at least 0.4 to form a bound system assuming plausible SN\,II explosion scenarios.
By analyzing the results we reveal that the orbit of a planetary companion is circularized in all cases.
An exception to the above statement occurs for the most massive neutron star models ($M_\mathrm{n}=3\,M_\odot$) with the least massive ejecta mass models ($M_\mathrm{ej}=6.5\,M_\odot$), shown in panel c3).
In this case, a bimodal peak can be found at $e=0.4$ and $e=0.9$, corresponding to $e_\mathrm{init}=0.8$ and $e_\mathrm{init}=0.4$, respectively, with the latter case meaning that the eccentricity of the planet increases.

Assuming a stellar-mass companion, the clumpy structure in eccentricity is absent, as shown in panels d1--d3 of Figure\,\ref{fig:hist_bound}.
We find that a bound system can form assuming arbitrary initial eccentricity (even a circular orbit with $e_\mathrm{init}=0$).
As a consequence both circularization and the excitation of eccentricity occur.
Stellar systems do not require a specific range of true anomalies to stay bound.
Contrary to the planetary-mass companion models, $e$ increases with increasing $M_\mathrm{sec}$.
A wider final stable orbit also coincides with a~more ~eccentric~one.

For stable binary systems, we observe that the system gains non-zero peculiar velocity, even though the explosion was assumed to be completely symmetric. 
For a planetary-mass secondary companion, this peculiar velocity is found to be on the order of $10^{-6} \, \mathrm{km/s}$, which is observationally insignificant.
To find the largest peculiar velocity binary, one must consider the smallest possible companion, highest ejecta mass, and eccentricity, and the secondary should be at pericenter, where the velocity of the components is the largest.
For stellar binaries, we find significantly higher velocities, up to $37 \, \mathrm{km/s}$, which can presumably be detected.

Investigation of the effect of ejecta velocity showed that higher $v_\mathrm{max}$ always results in a lower $e$ value for the bound system.
This finding is independent of the companions' mass, and thus is true for both planetary and stellar-mass companion models.

\subsection{Unbound cases}

\begin{figure*}
    \centering
    \includegraphics[width=0.75\textwidth]{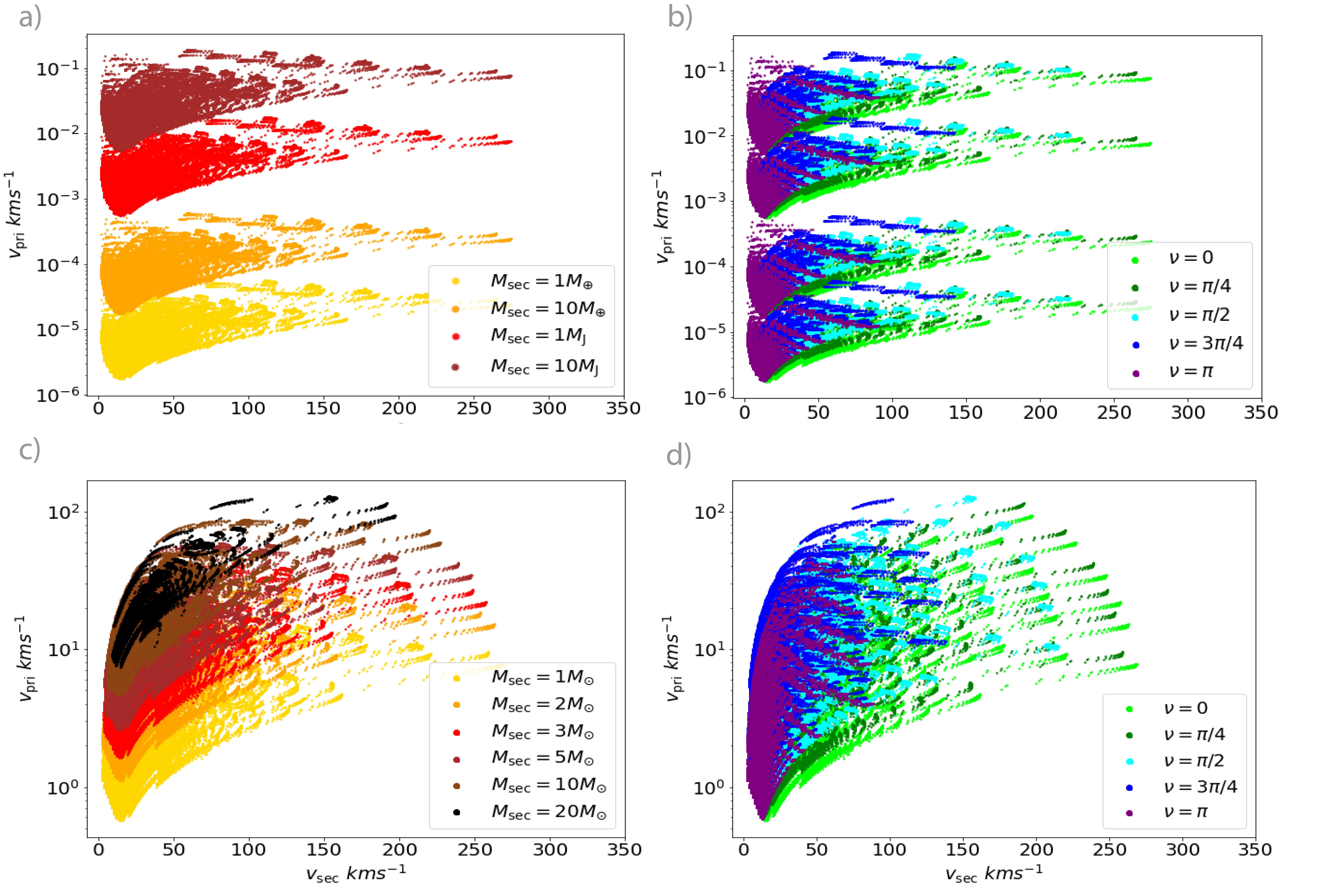}
    \caption{Distribution of the primary and secondary velocities of unbound planets (panels a and b) and stellar-mass companions (panels c and d) by the end of the simulation. Colors represent secondary mass on the left and initial true anomaly on the right panels. Values of $v_\mathrm{pri}$ are on a wide range and thus shown on a logarithmic scale, while $v_\mathrm{sec}$ values are on a linear scale.}
    \label{fig:velos_unbound}
\end{figure*}

\begin{figure*}
    \centering
    \includegraphics[width=0.9\textwidth]{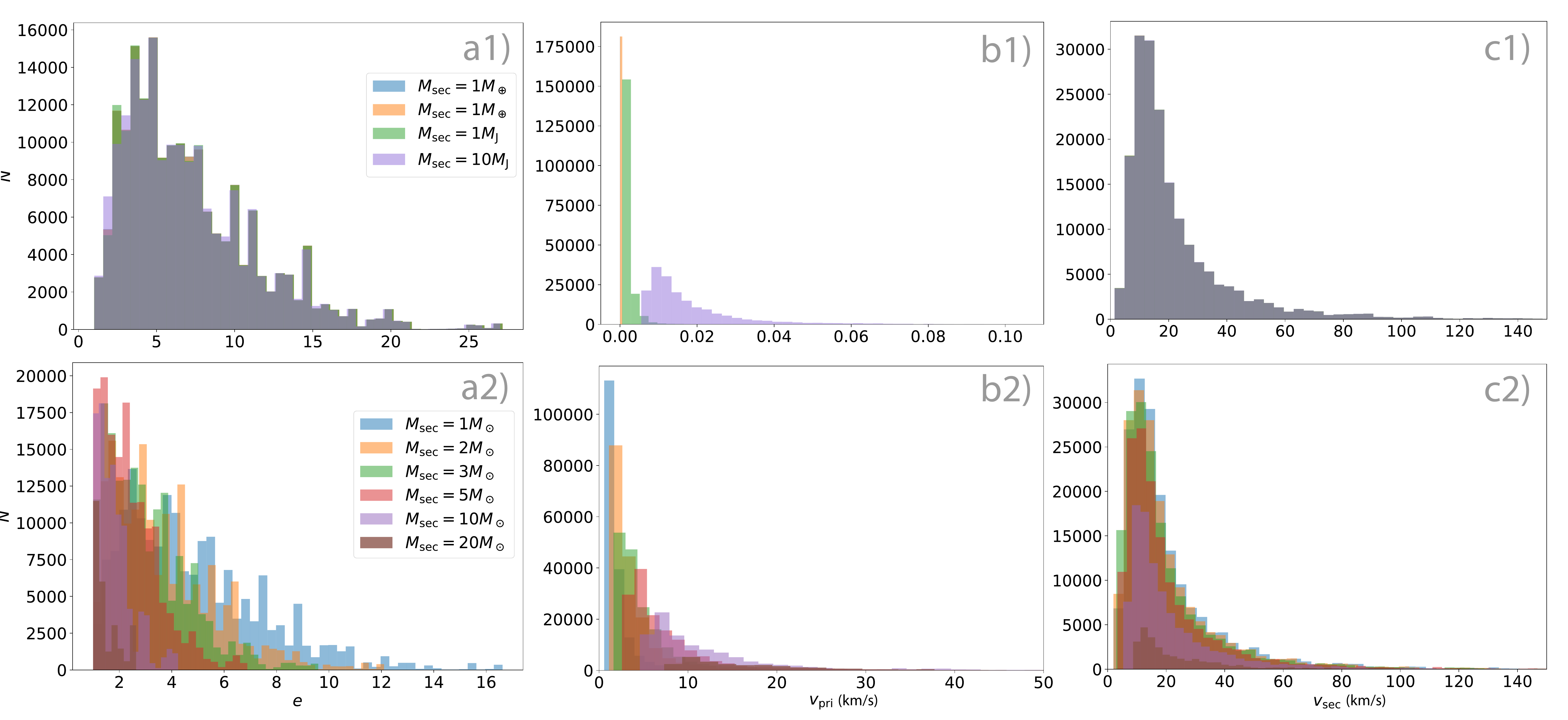}
    \caption{Histograms of the eccentricities and the final velocities of the neutron star and companion in the case where the planets (upper panels) and stars (lower panels) become unbound. Different colors represent different secondary masses, also indicated in the key. Different populations can be distinguished, whose width increases with $M_\mathrm{sec}$. The gray areas show completely overlapping regions.}
    \label{fig:hist_unbound}
\end{figure*}

For unbound systems, the final eccentricity of the secondary grows above unity, corresponding to a parabolic or hyperbolic unbound orbit.
In these scenarios, we analyze the distribution of the final velocities of both the primary and secondary components.
Figure \ref{fig:velos_unbound} shows $v_\mathrm{pri}$ and $v_\mathrm{sec}$ for an unbound state both for planetary and stellar companions. 
The histograms of $e$, $v_\mathrm{pri}$ and $v_\mathrm{sec}$ are also given in Figure \ref{fig:hist_unbound}.

For planetary-mass companion models, $v_\mathrm{pri}$ are grouped by $M_\mathrm{sec}$ and the velocity values increase with $M_\mathrm{sec}$; see panel a of Figure\,\ref{fig:velos_unbound}.
However, $v_\mathrm{sec}$ is independent of the mass of the planet.
Concerning the true anomaly, the sensitive parameter however is only the secondary's velocity.
One can see in panel b of Figure\,\ref{fig:velos_unbound} that the closer $\nu$ is to $\pi$ (i.e., apocenter), the lower the secondary velocity, while the primary velocity can span a wide range of values.

For stellar-mass companion models (panels c and d of Figure\,\ref{fig:velos_unbound}) $v_\mathrm{pri}$ values are three orders of magnitude higher than that of planetary-mass companion models, while $v_\mathrm{sec}$ values display the same range of values for both models.
Notably, low primary and high secondary velocity systems are absent, as are systems with high primary and low secondary velocities.
Both velocities are sensitive to stellar mass and true anomaly, as shown in panels c and d of Figure \ref{fig:velos_unbound}.
Based on the clear separation of colors representing the secondary's mass in panel c, a higher $M_\mathrm{sec}$ results in a higher $v_\mathrm{pri}$ and slightly lower $v_\mathrm{sec}$.
Similarly to planetary-mass companion models, low $v_\mathrm{sec}$ develops if the secondary is close to apocenter.

An interesting feature of the secondary eccentricities shown on panels a1 and a2 of Figure\,\ref{fig:hist_unbound} is that there exists a most likely value for planetary systems around $e=4$.
This trend is absent for stellar companions, where the larger the eccentricity the lower its occurrence.

Two populations of primary velocity (with narrow and wide distributions) can be distinguished; see panel b1 of Figure\,\ref{fig:hist_unbound}.
The width of the distribution of $v_\mathrm{pri}$ increases with $M_\mathrm{sec}$.
Very low primary velocities ( $\sim 10^{-6}-10^{-4}\,\mathrm{km\,s^{-1}}$) correspond to $M_\mathrm{sec}=1\,M_\oplus$ and $M_\mathrm{sec}=10\,M_\oplus$, while an order of magnitude larger primary velocities (up to $0.1\,\mathrm{km/s}$) correspond to $M_\mathrm{sec}=1\,M_\mathrm{J}$ and $M_\mathrm{sec}=10\,M_\mathrm{J}$.
The primary velocities for stellar companion models, shown in panel b2 of Figure\,\ref{fig:hist_unbound}, also exhibit distinct populations based on $M_\mathrm{sec}$, where a higher $M_\mathrm{sec}$ also yields a higher $v_\mathrm{pri}$ value.
The most likely value for primary velocity in the stellar case is $\sim 4\,\mathrm{km/s}$.

The secondary's velocity histograms of planetary and stellar companion models (panels c1 and c2 in Figure\,\ref{fig:hist_unbound}) are remarkably similar: the most likely secondary velocity is  $\sim18\,\mathrm{km/s}$, while the highest measured value is $\sim270\,\mathrm{km/s}$.
A notable feature of both primary and secondary velocities is that few models can result in velocities higher than $v_\mathrm{pri}=30\,\mathrm{km/s}$ and $v_\mathrm{sec}=\mathrm{100\,km/s}$ for a stellar companion.
For a planetary-mass companion, $v_\mathrm{pri}$ is smaller by three orders of magnitude.

By analyzing the primary and secondary velocities as a function of $v_\mathrm{max}$, we found no clear trend.
This means that neither the primary nor the secondary velocities are sensitive to the speed of the exploding envelope in the investigated parameter space.

\subsection{True anomaly}
\label{sub:true_anomaly}

In accordance with \citet{veras-etal}, we also found that in order to have a bound system, an eccentric orbit is required for planetary-mass companions.
For stellar companion models, however, a companion on a circular orbit initially can also remain bound to the neutron star.
For an eccentric orbit, only a certain range of initial true anomaly (true anomaly at the instance of explosion) values, $\nu_\mathrm{b}$, results in bound systems.

\begin{figure*}
    \centering
    \includegraphics[width=0.9\textwidth]{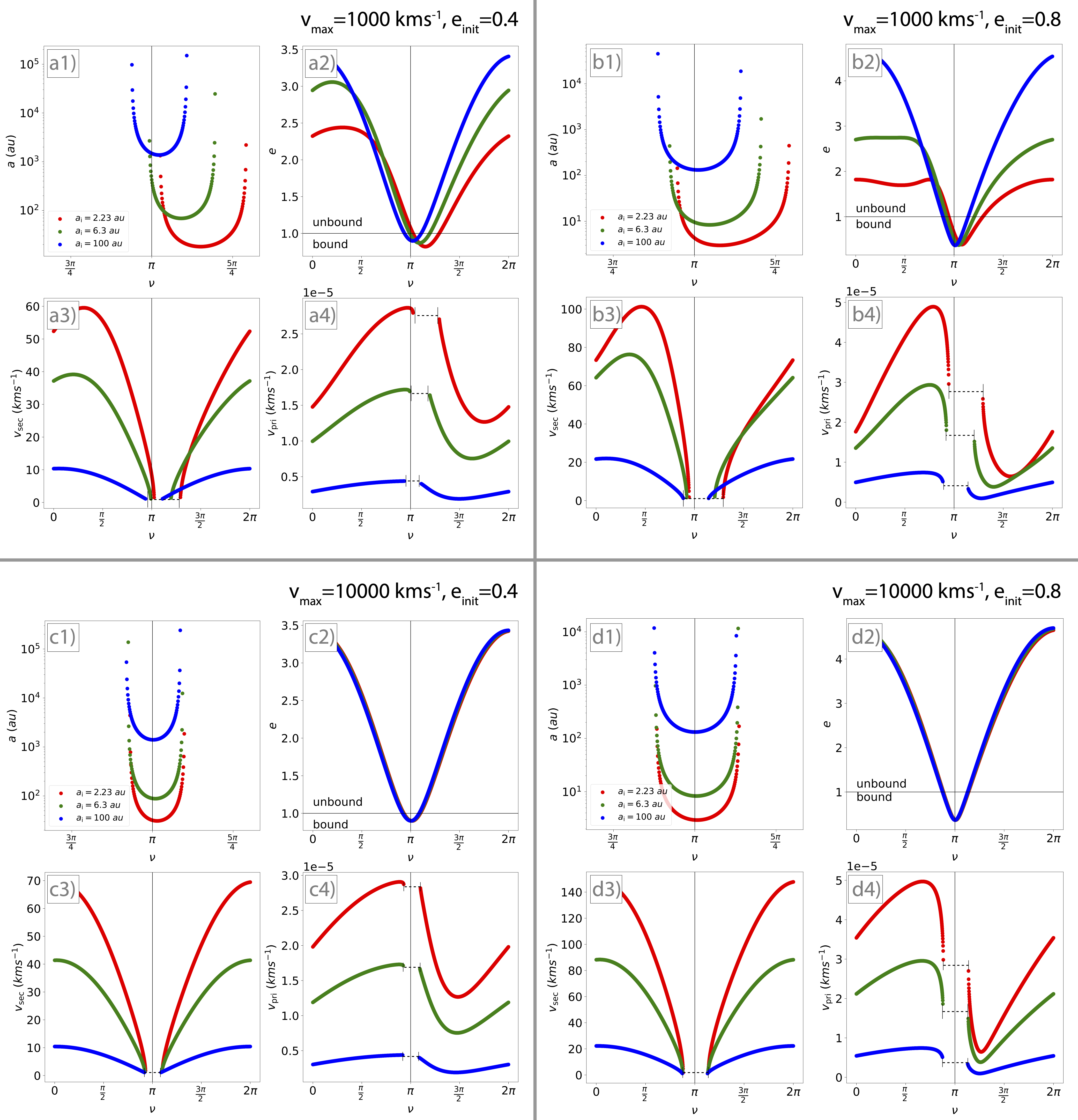}
    \caption{Groups of panels 1-4 show the final orbital parameters $a$, $e$, and component velocities as a function of $\nu$ assuming a $1\,M_\oplus$ secondary. The applied ejecta speed and initial eccentricity are given. Colors show different $a_\mathrm{init}$ values. Regions absent of data points (indicated with dashed lines) represent bound models on component velocity panels.} 
    \label{fig:true_anomaly_planet}
\end{figure*}

We thoroughly investigated the effect of true anomaly on planetary and binary systems with an initial eccentricity of either 0.4 or 0.8 and an ejecta speed of either 1000 or $10,000 \,\mathrm{km\,s^{-1}}$.
For the planet, we assumed $1\,M_\oplus$, though we find that the results can be generalized to the whole examined planetary mass range.
Results are shown in Figure\,\ref{fig:true_anomaly_planet} for three $a_\mathrm{init}$.
Four panels are shown for each model: $a$, $e$ , $v_\mathrm{pri}$, and $v_\mathrm{sec}$ as a function of $\nu$.

The higher the semi-major axis, the fewer systems stay bound in all model sets, as visible on panels a1, b1, c1, and d1 in Figure\, \ref{fig:true_anomaly_planet}.
It is clearly seen that $\nu_\mathrm{b}$ shifts toward higher initial true anomaly values if a larger $a_\mathrm{init}$ is used for low $v_\mathrm{max}$ models.
However, this trend diminishes for higher ejecta velocities; therefore $\nu_\mathrm{b}$ becomes symmetric around the apocenter.
As one can see in Figure \, \ref{fig:true_anomaly_planet}, for $v_\mathrm{max}=1000 \, \mathrm{km/s}$, $a_\mathrm{init}$ affects $e$, (panels a2 and b2); however, $e$ becomes independent of $a_\mathrm{init}$ for $v_\mathrm{max}=10,000 \, \mathrm{km\,s^{-1}}$, $e$ (see panels c2 and d2).
The width of the bound region is found to be proportional to $e_\mathrm{init}$ independent of $v_\mathrm{max}$ (meaning that more bound systems can form with higher $e_\mathrm{init}$).
If $v_\mathrm{max}=1000\,\mathrm{km/s}$, $\nu_\mathrm{b} \in [165^{\circ}-235^{\circ}]$, while for  $v_\mathrm{max}=10000\,\mathrm{km\,s^{-1}}$, $\nu_\mathrm{b} \in [155^{\circ}-200^{\circ}]$.
In general, a lower $a_\mathrm{init}$ results in lower eccentricities, clearly visible for unbound orbits.
However, with increasing $v_\mathrm{max}$, the eccentricity becomes independent of $a_\mathrm{init}$.

The primary and secondary velocities measured in unbound models show a strong dependence on $a_\mathrm{init}$ and $\nu$.
In general, the component velocities are found to decrease with  $a_\mathrm{init}$.
Interestingly, $v_\mathrm{sec}$ curves are highly symmetric around the apocenter for high $v_\mathrm{max}$ values, for which symmetry starts to break for a lower ejecta velocity.
Ny contrast, $v_\mathrm{pri}$ curves are never symmetric.
Note that, with $e_\mathrm{init}=0.8$, the velocities of components are almost doubled compared to $e_\mathrm{init}=0.8$, while $v_\mathrm{max}=1000\,\mathrm{km/s}$ results in less than 50\% variation in the secondary, and less than 10\% variation in primary velocity.
The above findings correspond quite well to the analytically derived true anomaly ranges in \citet{veras-etal}.
The asymmetry seen in Figure \ref{fig:true_anomaly_planet} can be explained by the difference in the definition of $\nu$; see details in Section \ref{sec:discussion}.

\begin{figure*}
    \centering
    \includegraphics[width=0.9\textwidth]{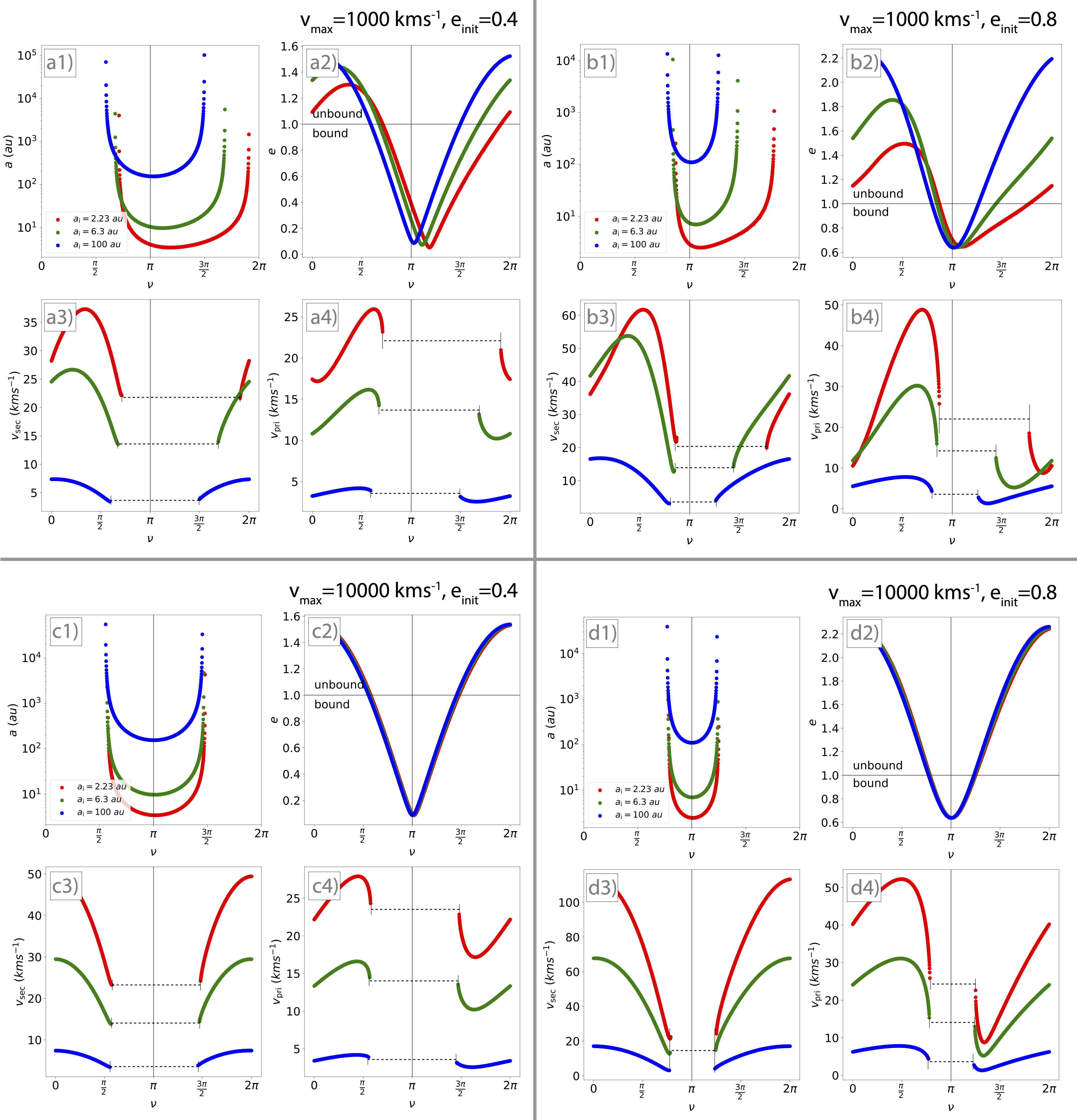}
    \caption{Same as Figure \ref{fig:true_anomaly_planet} for a stellar companion of $5\,M_\odot$.}
    \label{fig:true_anomaly_star}
\end{figure*}

Figure \ref{fig:true_anomaly_star} is similar to Figure \ref{fig:true_anomaly_planet}, the only difference being that we modeled a binary system in which the mass of the secondary is set to be $5\,M_\odot$.
One can see that $\nu_\mathrm{b}$ spans a wider range of values, $[110^\circ-340^\circ]$ (panel a3), but can be as low as $[140^\circ-225^\circ]$ (panel d3).
Moreover, it is clearly seen in the velocity panels (a3, a4, b3, b4, c3, c4, d3, and d4) that almost 50\% of models remain bound.
Concerning the symmetry and asymmetry of velocities, we found that the secondary velocities are highly asymmetric assuming $v_\mathrm{max}=1000\,\mathrm{km/s}$, contrary to the planetary case.
However, for $v_\mathrm{max}=10000\,\mathrm{km/s}$ the symmetry is somewhat restored.

Another important behavior of binary systems is that the number of bound systems is about two times higher for $e_\mathrm{init}=0.4$ than for $e_\mathrm{init}=0.8$, contrary to the planetary-mass companion models. 
This simply means that a binary system can remain bound if $e_\mathrm{init}$ is low. 
Finally, we found that all other conclusions drawn from the planetary companion models are also valid here.

\section{Discussion}
\label{sec:discussion}

\subsection{Comparing homologous and linear mass-loss}
\label{sub:comparison}

\begin{figure}
    \centering
    \includegraphics[width=\columnwidth]{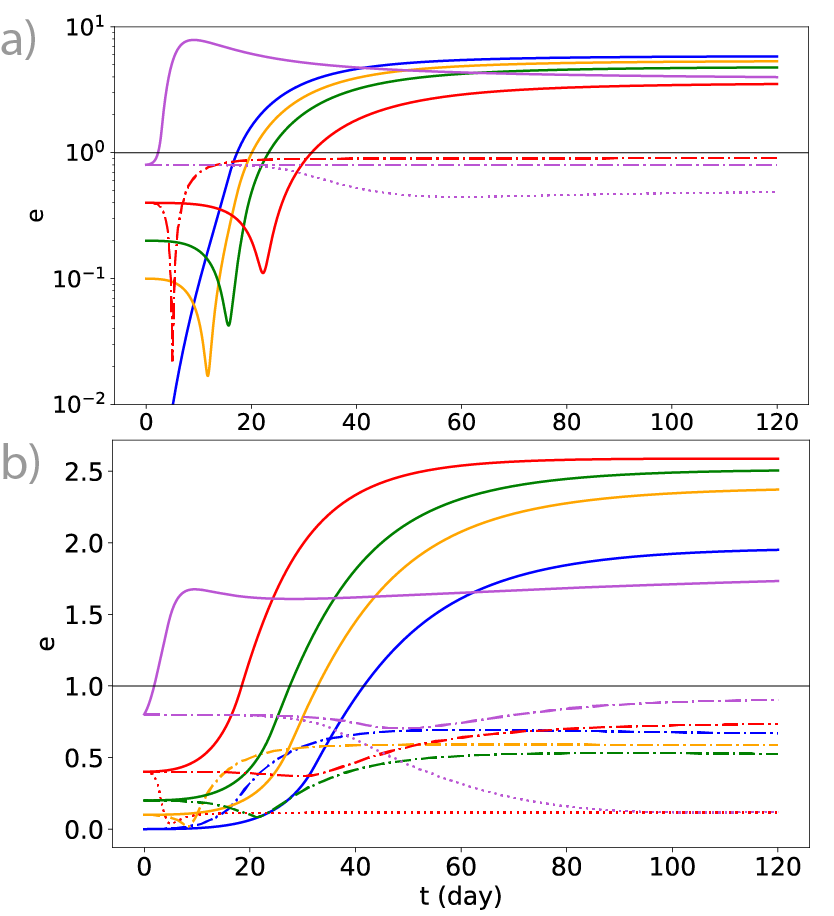}
    \caption{Representative eccentricity evolution models for planetary (a) and stellar (b) systems. Solid lines represent unbound systems, while dotted ($e<e_\mathrm{init}$) and dotted-dashed ($e>e_\mathrm{init}$) lines represent bound systems. Colors represent different $e_\mathrm{init}$ values, which are exactly the same for a given color. For visibility, we use a logarithmic scale in panel~(a).}
    \label{fig:e(t)}
\end{figure}

Figure~\ref{fig:e(t)} shows the evolution of the secondary's eccentricity as a function of time for all possible outcomes of models for both planetary (panel a) and stellar companions (panel b).
In agreement with \citet{veras-etal} we found that $e_\mathrm{init} \geq 0.4$ is required for the formation of a bound planetary system.
We also confirm that at early phases the eccentricity decreases, then attains a minimum value (which never reaches zero) and starts to increase for systems with $e_\mathrm{init} \geq 0.8$.

With regards to the stable orbits' dependence on true anomaly, $\nu_\mathrm{b}$ \citet{veras-etal} found that $\nu_\mathrm{b}$ is symmetric around the apocenter.
However, according to our findings $\nu_\mathrm{b}$ is asymmetric and is pushed toward higher $\nu$ values.
This contradiction can be resolved by the fact that the true anomaly is measured at the beginning of the explosion in our models.
If $\nu$ were measured at the instance when the outer envelope layer reaches the planet, $\nu_\mathrm{b}$ would be symmetric.
This means that asymmetry is simply caused by orbital movement of the secondary during the early phase of envelope expansion.

Our numerical approach made it possible to examine cases where the secondary mass is nonnegligible relative to that of the primary. 
With regard to these stellar-mass companion models (panel~b of Figure~\ref{fig:e(t)}), we got novel results.
Although bound systems show similar behavior to planetary models, decreasing eccentricity is not observed for free-floating stars. 
Another major difference compared to planetary systems is that there is no minimum requirement of $e_\mathrm{init}$ to form a bound orbit.

\begin{figure*}
    \centering
    \includegraphics[width=0.75\textwidth]{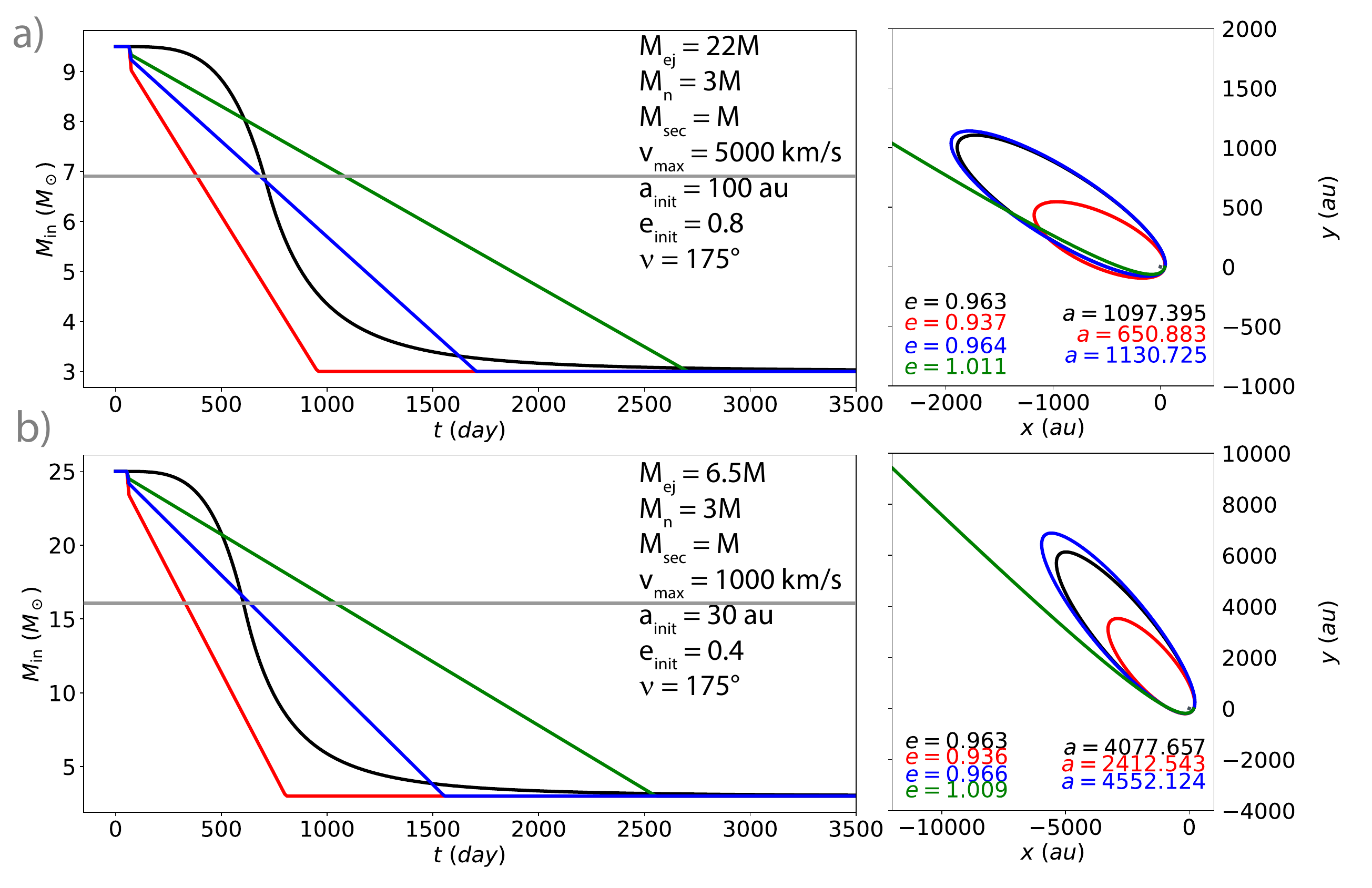}
    \caption{Mass loss as a function of time and orbital parameters by the end of the integration using the homologous expansion model (loss function and orbit in black). Colored lines show different linear mass-loss regimes. The secondary is a $1\,M_\oplus$ planet. The gray line marks 60\% of the ejecta mass, used for the constant mass-loss approximation.}
    \label{fig:diff_planet}
\end{figure*}

In \citet{veras-etal} a simple linear mass-loss model was assumed, in which case $\dot{M_\mathrm{in}}$ is constant.
In our homologous expansion model described in Section \ref{sec:model}, we apply a more elaborate version of an explosion model, in which the magnitude of $\dot{M_\mathrm{in}}$ is not time-invariant (see Figure \ref{fig:density}).
In this case $\dot{M_\mathrm{in}}$ slowly increases then gains a constant value, which slowly saturates to zero at later stages.
To compare the two explosion models, we conduct a set of representative planetary and stellar-mass companion models.
Three linear models are compared with different mass-loss rates in each case.
Figure \ref{fig:diff_planet} shows two planetary-mass companion models assuming $v_\mathrm{max}=5000\,\mathrm{km/s}$ (panel a) and $v_\mathrm{max}=1000\,\mathrm{km/s}$ (panel b).
In both runs, we found that the linear models can give qualitatively and quantitatively different results compared to the homologous model.
If the constant mass-loss rate is slightly smaller than in the homologous model, then the semi-major axis of the bound state can be smaller.
However, if the mass-loss rate underestimates that of the homologous expansion, the system can become unbound.
The linear model departs from the homologous model if it assumes the constant mass-loss rate value of the homologous model.
The observed differences hold regardless of the ejecta speed.
We found an easy method to get a somewhat similar solution for constant and homologous mass-loss models: appropriate $\dot{M_\mathrm{in}}$ can be derived by fitting a straight line connecting the primary mass and $\sim0.6\,M_\mathrm{ej}$ points.

For stellar-mass companion models, the homologous and linear models give qualitatively the same results. 
Variation in $\dot{M_\mathrm{in}}$ of the linear model, however, can give slightly different orbital parameters in bound states, or different velocity vectors in unbound states.

\subsection{Caveats and limitations}

Although we use an elaborate mass-loss approximation, the homologous expansion model intrinsically assumes spherical symmetry.
This simplification cannot explain most of the observations (e.g. planet-hosting PSR B1257+12's peculiar velocity of $326\,\mathrm{km/s}$ (see \citet{yan-etal}, or the velocities presented in \citet{hobbs-etal}).
A pulsar with a high peculiar velocity (e.g. $\sim1100\,\mathrm{km\,s^{-1}}$ in the case of  PSR B1508+55; \citealp{gvaramadze-gualandris}) can be indirect proof of a strongly asymmetric explosion.
Asymmetric envelope ejection complicates the perturbation of the orbital elements of the secondary, as well as the change in orbital elements of the primary (see, e.g., \citealp{parriott-alcock, namouni, namouni-zhou}).
In this case, our prediction for $v_\mathrm{pri}$ and $v_\mathrm{sec}$ are presumably inadequate.

With regard to the envelope size, $R_0$, we assumed it to be independent of primary mass, which predicts certain $M_\mathrm{ej}$--$M_\mathrm{n}$ pairings, which might affect the results.

By the nature of a one-dimensional approximation, the secondary-envelope interactions and the subsequent orbital perturbations are also neglected in this study.
For example, local perturbation in the envelope mass distribution by the secondary can break the assumed spherical symmetry.
As a result, both the secondary and primary orbital elements are subject to change.
Moreover, in a three-dimensional model, an additional orbital element, i.e., inclination, should be taken into account, which further complicates the picture, though the numerical N-body simulation for this limited number of bodies is not computationally demanding.

In a more sophisticated model, the secondary-envelope interactions (see, e.g.,  \citealp{debes-sigurdsson}) could result in mass-loss of the secondary, which effect was completely neglected in our orbital solutions.
To investigate this phenomenon, a mass loss and accretion model for the secondary is required.
Presumably, this effect is more important for stellar binary configurations.

Handling the abovementioned limitations may affect the orbital parameters of bound systems and velocity components of unbound systems, or can even give qualitatively different results.
Future investigations will also address subsequent supernova explosions in binary systems and hierarchical triple system configurations (circumbinary -- continuing the work of \citealp{fagginger-portegies}, circumprimary, circumsecondary planets, and star-giant~planet-moon systems).
These hierarchical systems can be investigated with our homologous expansion model.
Adding another planet \citep{debes-sigurdsson} and adding a Kuiper belt \citep{bonsor-etal}
have been investigated in detail for white dwarfs and a white dwarf with a main-sequence companion.
We emphasize that both studies use extremely low mass-loss rates ($10^{-8}$--$10^{-6}\,M_\odot\,\mathrm{yr^{-1}}$, $2/3\times 10^{-5}\,M_\odot\,\mathrm{yr^{-1}}$), which resemble asymptotic giant branch mass loss.

\subsection{Pulsar planets and rogue planets}

Our results imply that the formation of close-in ($a_\mathrm{init} < 2\,\mathrm{au}$) pulsar planets requires additional physics, as bound systems have, in general, much wider separation. 
Note that PSR B0329+54 b \citep{starovoit-rodin} can be explained by our simple model if the system was initially in a high eccentricity state ($e_\mathrm{init} \geq 0.4$) and the planet was very close to the stellar surface.
Since the final-to-initial semi-major axis ratio is found (in the investigated parameter space) to be always at least 2, pulsar planets should have orbited the progenitor inside half of their current semi-major axis.
We emphasize that a bound system also requires fortunate explosion timing, i.e., the planet should be close to its apocenter.

With regards to the first exoplanetary system discovered around a pulsar, PSR B1257+12 \citep{wolszczan-frail, wolszczan-1994, wolszczan-2012}, we think that the probability of capturing a single planet is low, and multiple captures are highly improbable.
The missing physics to explain the existence of such systems could be a second-generation planet formation hypothesis, in which the pulsar planets form after the explosion from the fallback material.
Although disks formed this way would be highly irradiated by the PSR \citep{martin-etal}, they are found to be similar in mass to protoplanetary disks around low-mass young stars \citep{lin-etal, menou-etal, williams-cieza}.
Planet formation scenarios in PSR disks have been studied and summarized in \citet{phinney-hansen}, \citet{thorsett-etal}, and \citet{podsiadlowski}.

An interesting consequence of our results is that rogue planets most probably originate from a low eccentricity orbit.
These planets have an average velocity of $1-275\,\mathrm{km/s}$.
In these models, the newborn pulsar gains a peculiar velocity regardless of the symmetric nature of the explosion.
Therefore, low peculiar velocity (up to $0.1\,\mathrm{km/s}$ for Jupiter-sized, and $10^{-3}\,\mathrm{km/s}$ for Earth-sized planets) pulsars most probably originate from pulsar-planet systems.

\subsection{Remnant binaries and free-floating stellar corpses}

Bound stellar systems can later become neutron star--white dwarf pairs or binary neutron stars depending on secondary mass.
Simulations in this study reveal that the eccentricity and semi-major axes of these systems show a homogeneous distribution on a wide range.
Post-explosion binary systems will gain a peculiar speed of up to $37\,\mathrm{km\,{s^-1}}$ in the investigated parameter space.
This can be explained by the fact that the exploding shell inherits the velocity of the primary at the moment of explosion, while the binary gains equal but opposite momentum due to the conservation of momentum.
This can be a novel channel for the formation of young expanding shells missing the central pulsar, besides the obvious explanations of the nondetection of the pulsar signal or an asymmetric explosion.
Another possible outcome can be the production of peculiar speed (a few tens of km/s) binary stellar corpses after a second explosion.
The formation of binary neutron star systems is going to be the subject of a future study, where a subsequent explosion of the secondary will be modeled.
Based on our results we predict that the binary neutron star velocity can be two times as high as observed in the single-explosion models, which is still below $100 \, \mathrm{km\,s^{-1}}$.

Disruption of binary systems can produce two groups of stellar remnants.
In the first group, there is a pulsar and a white dwarf, for which case the progenitor of the white dwarf is under the mass limit of an SN~II explosion; it is $\leq 8\,M_\odot$.
In this case, our results show that the most likely primary pulsar velocities are $1-25\mathrm{km\,s^{-1}}$, while white dwarf velocities are $1-128\,\mathrm{km\,s^{-1}}$.

In the second group, two neutron stars are born, and the primary neutron star velocity is most likely in the range of $5-40\,\mathrm{km\,s^{-1}}$, while the velocity maximum of the secondary is decreased to $100\,\mathrm{km\,s^{-1}}$.
The increased velocity of the primary neutron star is due to the fact that the secondary should have a mass above $8\,M_\odot$. 

We found an upper limit of neutron star velocity ($275\,\mathrm{km/s}$) based on our secondary velocity measurements.
This prediction implicitly assumes that the secondary's subsequent SN~II explosion is fully symmetric, so does not change the velocity vector of the neutron star.
Contradicting our results, PSR B1508+55 has a ~1100 \, km/s peculiar velocity \citep{gvaramadze-gualandris}; however the assumption of a strongly asymmetric explosion might resolve this inconsistency.

\section{Conclusions}
\label{sec:conclusions}

In this paper, we studied the fate of planet-star and binary star systems undergoing a supernova explosion of the primary component.
Our study utilizes over two and a half million numerical simulations.
We gave a simplified model for homologous envelope loss, which was combined with a high-precision N-body solver.
This study continues the pioneering work of \citet{veras-etal} by investigating scenarios in which the secondary mass is comeasurable to that of the primary.
We investigated systems in which the primary mass is larger than an SN\,II explosion's lower limit.
Our investigation is limited to an ejecta velocity greater than $1000\,\mathrm{km/s}$ to model envelope ejection in SN~II explosions.
Due to the nature of the applied numerical methods, we were able to calculate the change in orbital elements of the secondary, as well as the velocity components of systems that fall apart. 
Our major findings are the following:

1) We confirmed the findings of \citet{veras-etal} with a homologous expansion model: a narrow range of true anomaly centered at the apocenter and a high initial eccentricity ($e_\mathrm{init} \geq 0.4$) are required to form a bound pulsar-planet system.
Bound systems can gain a large semi-major axis (several thousand au) and high eccentricity ($>0.9$) in accordance with \citet{thorsett-etal}.

2) We identified two scenarios in bound pulsar-planet models in which the planetary eccentricity increases ($e_\mathrm{init}=0.4$) or decreases ($e_\mathrm{init}=0.8$). Furthermore, a higher ejecta speed is found to result in a lower eccentricity for the planet.

3) Concerning planetary systems that fall apart, we found that the neutron star velocity is $10^{-6}-10^{-1}\,\mathrm{km\,s^{-1}}$ and is proportional to $M_\mathrm{sec}$.
The planet velocity is $1-275\,\mathrm{km\,s^{-1}}$, with a most likely value of $18\,\mathrm{km\,s^{-1}}$ independent of its mass.
There is a maximum velocity of free-floating planets as a function of $\nu$, which does not coincide with $\nu=\pi$ but is rather shifted to slightly larger values for $v_\mathrm{max}=1000\, \mathrm{km\,s^{-1}}$.
For $v_\mathrm{max}=10,000\, \mathrm{km\,s^{-1}}$, this maximum coincides with $\nu=\pi$.

4) We also investigated the dependence on the initial true anomaly, upon which the region of stability, eccentricity, and velocities are dependent. At low ejecta speeds, higher $a_\mathrm{init}$ results in higher final eccentricity. For unbound systems, eccentricity values are less than $4.5$.

5) For stellar systems that fall apart, the primary velocity is found to be three orders of magnitude higher ($1-100\,\mathrm{km\,s^{-1}}$) than those of planetary systems, which also shows a clear dependency on $M_\mathrm{sec}$.
However, $v_\mathrm{sec}$ falls in the same range for stellar-mass secondary systems as it does for planetary-mass secondary systems ($1-275\,\mathrm{km\,s^{-1}}$).
The same trend in $v_\mathrm{sec}$ can be seen as for planetary systems regarding true anomaly.
There also exists a most likely speed for unbound neutron stars, which is around $v_\mathrm{pri}=4\,\mathrm{km,s^{-1}}$, and for unbound secondaries around $v_\mathrm{sec}=18\,\mathrm{km\,s^{-1}}$.

6) Investigating the effect of true anomaly on binary star systems, we found that for low ejecta speeds, a higher $a_\mathrm{init}$ yields a higher $e$ value, which does not go over 2.2 even for unbound stars.

7) Finally, we gave a simple method to derive an approximated constant mass-loss rate from a homologous model, which can give a qualitatively and quantitatively similar result to the homologous model.

Assuming a homogeneous sampling of initial parameters results in one-and-a-half orders of magnitude more free-floating or so-called rogue planets that can form than bound systems in SN\,II explosion. 
This result is in accordance with the findings of \citet{kerr-etal} and \citet{manchester-etal} that very few planet-hosting pulsars are observed.
So far, rogue planets have been thought to originate from scaled-down core collapse \citep{padoan-nordlund, hennebelle-chabrier}, later instabilities and ejection \citep{veras-raymond}, aborted stellar embryos \citep{reipurth-clarke}, and photoerosion of prestellar cores \citep{whitworth-zinnecker}.
All of these processes work to some degree \citep{testi-etal, fontanive-etal}, but still our results present a novel channel of rogue planet formation: SN~II explosions.

Finally, let us emphasize some of our main conclusions based on our symmetric SN~II explosion models.
We think that the most plausible explanation for close-in pulsar planets (e.g. PSR~B1257+12, PSR~J1719-1438, PSR~B0943+10, and PSR~J2322-2650) is post-explosion planet formation because the semi-major axis at least doubles during mass loss.
Rogue planets can gain up to $275\,\mathrm{km\,s^{-1}}$ peculiar velocities in an SN~II explosion scenario, which can also result in low velocity ($<0.1\,\mathrm{km\,s^{-1}}$) free-floating pulsars. 
Interestingly, this SN~II explosion scenario could produce high peculiar velocity binaries of stellar corpses with a velocity of less than $100\,\mathrm{km\,s^{-1}}$.
We also conclude that pulsars discovered with peculiar velocities of less than $25\,\mathrm{km\,s^{-1}}$ or between 25 and $275\,\mathrm{km\,s^{-1}}$ can form from the primary or the secondary undergoing an SN~II explosion, respectively.
White dwarfs discovered with a peculiar velocity of less than $128\,\mathrm{km\,s^{-1}}$ may have also formed in the same scenario, assuming that the secondary is under the SN~II mass limit.

\acknowledgments {
The project is supported by the project "Transient Astrophysical Objects" GINOP 2.3.2-15-2016-00033 of the National Research, Development and Innovation Office (NKFIH), Hungary, funded by the European Union, and by the NKFIH OTKA Grant K-142534.
}

\end{document}